\documentclass[journal=jpcl,manuscript=letter]{achemso}

\usepackage{chemformula} 
\usepackage[T1]{fontenc} 
\usepackage{amsmath,upgreek,xcolor,braket}
\usepackage{hyperref}
\usepackage{microtype}

\author{\normalsize Anne M. Fabricant}
\affiliation{Department of Biosignals, Physikalisch-Technische Bundesanstalt (PTB), 10587 Berlin, Germany}
\alsoaffiliation{Institute of Physics, Johannes Gutenberg University of Mainz, 55099 Mainz, Germany}
\alsoaffiliation{Helmholtz Institute Mainz, 55099 Mainz, Germany}
\author{Florin Teleanu}
\affiliation{Department of Chemistry, New York University, New York, NY 10003, United States}
\alsoaffiliation{Extreme Light Infrastructure - Nuclear Physics, ``Horia Hulubei'' National Institute for Physics and Nuclear Engineering, 077125 Bucharest, Romania}
\author{Adam Ortmeier}
\affiliation{Department of Chemistry, North Carolina State University, Raleigh, NC 27607, USA}
\author{Piotr Put}
\affiliation{Faculty of Physics, Astronomy and Applied Computer Science, Jagiellonian University in
Kraków, 30-348 Kraków, Poland}
\author{Seth Dilday}
\affiliation{Department of Chemistry, North Carolina State University, Raleigh, NC 27607, USA}
\author{Alexej Jerschow}
\affiliation{Department of Chemistry, New York University, New York, NY 10003, United States}
\author{Dmitry Budker}
\affiliation{Institute of Physics, Johannes Gutenberg University of Mainz, 55099 Mainz, Germany}
\alsoaffiliation{Helmholtz Institute Mainz, 55099 Mainz, Germany}
\alsoaffiliation{GSI Helmholtzzentrum f\"ur Schwerionenforschung, 64291 Darmstadt, Germany}
\alsoaffiliation{Department of Physics, University of California, Berkeley, CA 94720, USA}
\author{Thomas Theis}
\affiliation{Department of Chemistry, North Carolina State University, Raleigh, NC 27607, USA}
\alsoaffiliation{Current address: Department of Chemistry, University of California, Davis, CA 95616; UC Davis Department of Radiology, Sacramento, CA 95817, USA}
\email{ttheis@ucdavis.edu}
\author{Danila Barskiy}
\affiliation{Institute of Physics, Johannes Gutenberg University of Mainz, 55099 Mainz, Germany}
\alsoaffiliation{Helmholtz Institute Mainz, 55099 Mainz, Germany}
\alsoaffiliation{GSI Helmholtzzentrum f\"ur Schwerionenforschung, 64291 Darmstadt, Germany}
\email{barskiy@miami.edu}
\altaffiliation{Current address: Frost Institute for Chemistry and Molecular Science, Department of Chemistry, Department of Physics, University of Miami, Coral Gables, FL 33146, USA}

\title[]
  {\Large Quadrature-detected zero-to-ultralow-field nuclear magnetic resonance
  }

\abbreviations{NMR,ZULF}

\begin{document}

\begin{tocentry}

\centering
\includegraphics[height=4.6cm]{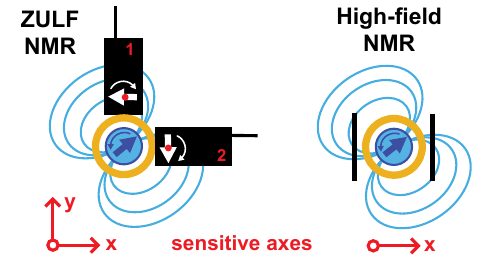}

Unlike in high-field NMR, quadrature detection in zero-to ultralow-field (ZULF) NMR measures the absolute direction of magnetization precession, providing a path to distinguishing signs of gyromagnetic ratios as well as simplifying spectral analysis. 

\end{tocentry}

\clearpage
\begin{abstract}

We present the concept of quadrature-detected zero-to-ultralow-field nuclear magnetic resonance (QD ZULF NMR). 
Two orthogonal sensors, atomic optically pumped magnetometers, are used to separately detect components of oscillating sample magnetization, which are then combined into a quadrature signal. This approach enables the distinction of positive and negative resonance frequencies with respect to the laboratory frame---such that absolute signs of gyromagnetic ratios can be determined. 
In addition, QD ZULF NMR reduces peak density and overlap compared to ZULF NMR with single-axis detection, while providing richer information about nuclear-spin dynamics. 
As an illustration, we present experimental $J$-spectra for thermally polarized [$^{13}$C]-formic acid and [$^{13}$C]-methanol, compounds commonly used in ZULF NMR which 
represent XA and XA$_3$ spin systems, respectively. A theoretical framework describing QD ZULF NMR signals is provided.

\end{abstract}



Quadrature detection (QD) is a concept widely encountered in signal acquisition across disciplines---from radio and television reception to various forms of spectroscopy. It refers to acquiring a signal consisting of two components that oscillate 90$^{\circ}$ out of phase (``in quadrature’’), enabling improved signal fidelity and the ability to gain more information about the signal. 
This is typically accomplished by specific acquisition hardware in combination with data-processing techniques. 
In nuclear magnetic resonance (NMR), QD involves extracting both orthogonal components of a spin magnetization vector, usually in the rotating frame of a given reference frequency~\cite{Levitt2008,Keeler2010}. Although this concept is well-established in conventional high-field and benchtop NMR, the analogous  realization has not yet been demonstrated in ZULF NMR.

In conventional NMR, excitation and detection are referenced to a carrier (receiver) frequency, such that only \textit{relative} resonance frequencies are measured. By contrast, in zero-to-ultralow-field (ZULF) NMR~\cite{Ledbetter2011,Theis2013,BlanchardBudker2016,Blanchard2021,Barskiy2025}, 
signal components are referenced to zero frequency (laboratory  frame), and hence \textit{absolute} reference frequencies are measured. In either case, a single-axis vector sensor---whether inductive or optical---is not generally capable of acquiring 
information about the direction of magnetization precession, unless initial magnetization is prepared orthogonally to the readout axis. 

Absolute frequency information is typically inaccessible at high field because only a cosine-modulated signal is recorded in each detector. This signal is 
split into real and imaginary parts, both of which are then mixed with the carrier frequency. Finally, a low-pass filter is applied such that only differences from the carrier frequency are stored. Furthermore, the excitation coil in high-field NMR excites just a narrowband spectral region. For example, in a 9.4\,T magnet, $^1$H spins resonate at 400\,MHz. This requires a pulse modulated at 400 MHz, which has the form $B_p\left(t\right)=B_p\left(0\right)\cos\left(2\pi\cdot400\,\text{MHz}\cdot t\right)$ in the time domain. Because of the cosine modulation, the pulse excites both the 400\,MHz and the $-400$\,MHz spectral regions, such that spectra are symmetric with respect to absolute zero, and frequency information about the sign of $\gamma_{\text{H}}$ is lost~\cite{Levitt1997}. We note that spectral information on either side of the carrier frequency is distinguishable, within the detection bandwidth and chemical-shift resolution, through the use of quadrature detection at high field. However, specially designed hardware (e.g., the use of birdcage coils) is needed to preserve absolute frequency information \cite{Doty1999}.

The fact that ZULF NMR intrinsically allows direct access to absolute resonance frequencies opens the possibility of distinguishing positive and negative spectral components with an appropriate orientation of the detector. Such frequency discrimination is critical for determining the relative signs of  gyromagnetic ratios ($\gamma$) and scalar couplings ($J$), 
and for accurately probing the symmetry of spin-state evolution~\cite{King2017,Georgiou2024,Wadhwa2024}. 
Despite this potential advantage, most ZULF-NMR systems to date have not been constructed to determine the absolute sign of resonance frequencies. Even with the correct choice of geometry, technical limitations (such as the difficulty of reconstructing the initial phase of a single-channel ZULF signal) have made separate measurement of orthogonal signal components desirable. This can be accomplished using two sensors oriented orthogonally in space, or by using a single sensor with dual-axis sensitivity~\cite{Osborne2018}. In this work, we combine these two approaches into a gradiometric QD scheme, which combines gradiometry for noise reduction with QD for frequency discrimination.

\begin{figure}
  \includegraphics[trim=150pt 0pt 150pt 0pt, clip,width=\textwidth]{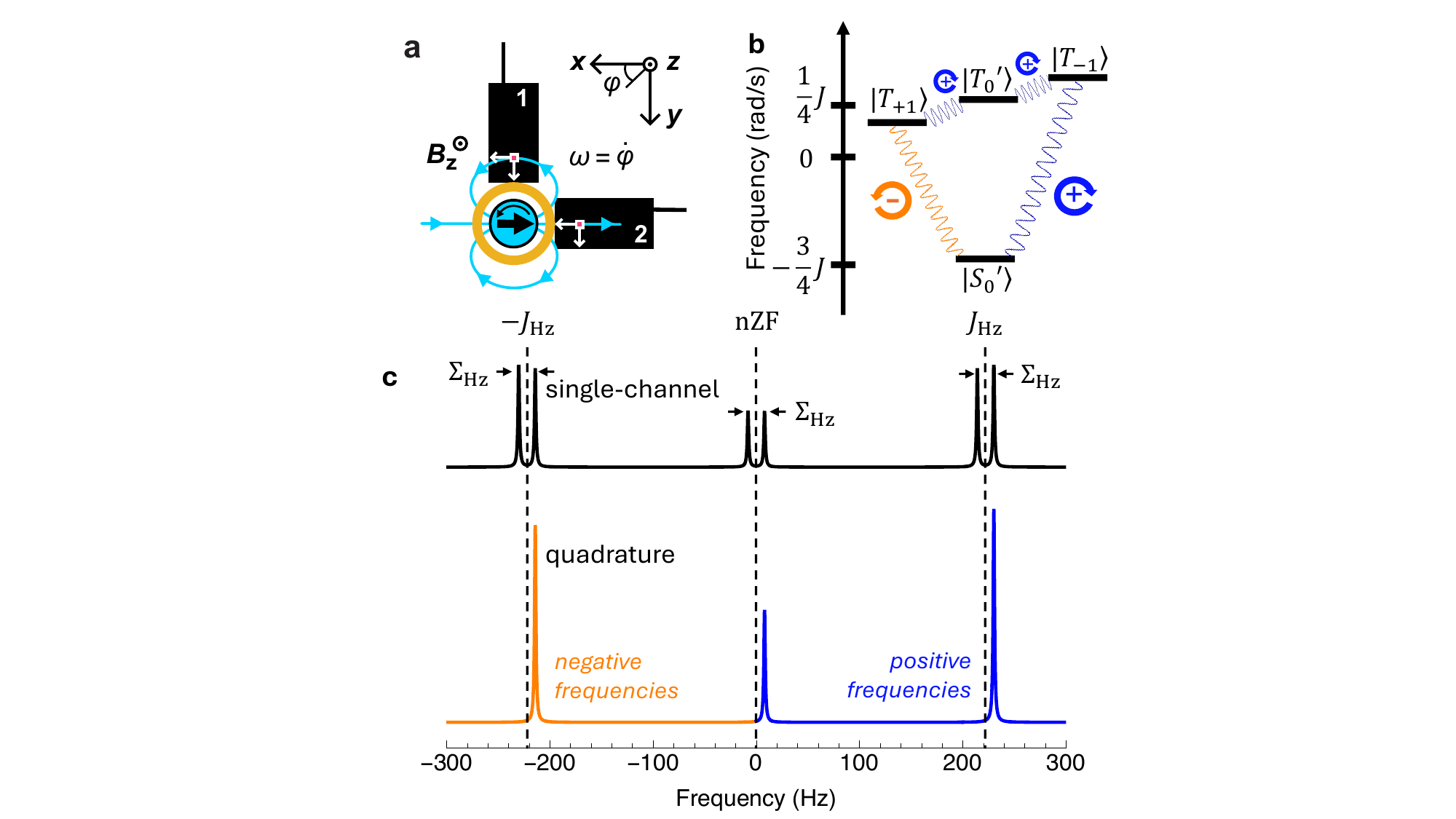}
  \caption{
  Gradiometric quadrature-detection (QD) 
  scheme for ZULF NMR~\cite{Fabricant2024,Fabricant2026}. (a) 
  Spin magnetization oscillating in the $x$-$y$ plane at angular frequency $\omega$ produces a magnetic field at the location of two magnetometer sensors, 1 and 2, placed orthogonally at equal distance from the sample. A solenoid (yellow circle) enables targeted application of a magnetic field $B_z$ to the sample and not the sensors. Each magnetometer is sensitive to fields along two orthogonal axes in the $x$-$y$ plane, as indicated by white arrows.
  For the case of magnetization precession, the rotation direction depends on the sign of the gyromagnetic ratio $\gamma$ as well as the sign of $B_z$---such that a $+z$-field induces counterclockwise precession for spin systems with $\gamma<0$.
  If the magnetization produces an effective dipole magnetic field, the maximum field amplitude of the measured radial channels ($x_2$, $y_1$) is twice that of the transverse channels ($x_1$, $y_2$).
  (b) Energy-level diagram of a scalar-coupled heteronuclear two-spin-1/2 system in the presence of the perturbing bias field $B_z$, with transitions detectable along the $x$- and $y$-axes transverse to the field.
  (c) Simulated magnitude spectra comparing single-axis (top) and quadrature (bottom) detection, as explained in the text, highlighting the 2-fold increase in signal of the latter.
  Gradiometric signals constructed by subtracting the pairs of time-dependent signals along $\hat{x}$ and $\hat{y}$, respectively, are combined into a complex function to obtain the quadrature signal. By performing a Fourier transform, signals of positive and negative frequency become distinguishable in the resulting spectrum. 
  } 
  \label{fig:quad}
\end{figure}

Two atomic magnetometers, each with dual-axis sensitivity, are placed orthogonally in the $x$-$y$ plane to detect spin-signal components oscillating at low frequencies ($<500$\,Hz) in the laboratory frame (Fig.\,\ref{fig:quad}a)~\cite{Fabricant2024,Fabricant2026}. At zero field, NMR signals of small organic molecules are typically composed of Zeeman peaks (which we also call near-zero-frequency, nZF, peaks) 
and oscillating magnetization due to $J$-coupling interactions ($J$-peaks, usually $<300$~Hz)~\cite{Theis2013,Put2021,Barskiy2025}. Upon application of a perturbing bias field, the energy levels of the spin system shift, resulting in frequency shifts and splitting of nZF peaks and $J$-peaks (Fig.\,\ref{fig:quad}b). 
More precisely, the selection rules for transitions are $\Delta F = 0, \pm 1$ and $\Delta m_{F} = 0, \pm 1$ \cite{Ledbetter2009,Ledbetter2011,Butler2013,Theis2013}.  A $\Delta m_{F} = 0$ transition corresponds to a driving field or detected magnetization oscillating along $z$, while $\Delta m_{F} = \pm 1$ transitions correspond to (counter-)rotating transverse fields or magnetizations \cite{Barskiy2025}.
The quadrature arrangement allows construction of a complex signal from the individually measured components, enabling phase-sensitive detection and spectral separation of positive- and negative-frequency resonance lines (Fig.\,\ref{fig:quad}c and Fig.\,S1).


In order to gain intuition for QD ZULF NMR and $J$-spectroscopy, we begin our discussion by reviewing the general mathematical framework of complex-signal detection suitable for explaining experimentally observed signals. Assuming a magnetization transverse to the detection field $B_z$ and aligned with a single-axis detector, when an oscillating signal is recorded at a frequency amplitude $\omega_0 =  -\gamma B_{z} $, the in-phase component is cosine-modulated according to $\cos \left(\omega_0 t\right) = \left( e^{i \omega_0 t} + e^{-i \omega_0 t} \right) / 2$ while the out-of-phase component is sine-modulated as $\sin \left(\omega_0 t\right) = \left( e^{i \omega_0 t} - e^{-i \omega_0 t} \right) / (2 i )$. In the absence of complete phase information, therefore, single harmonic modulation does not discriminate between negative and positive frequencies. 
However, the combination of the two measurements enables construction of a complex quadrature signal, defined as
\begin{equation}
\label{eq:signal}
\cos (\omega_0 t ) + i \sin (\omega_0 t ) = e^{i \omega_0 t} \,.
\end{equation}
This signal now offers vector information about the oscillation frequency $\omega_0$, and, given that $\omega_0$ is an absolute frequency, also reveals the absolute sign of $\gamma$ with respect to a known reference. 

First, we consider the simplest case of an XA (two-spin) heteronuclear system. This is typically encountered in $^{13}$C-labeled small molecules like [$^{13}$C]-formic acid, where the spin-zero $^{12}$C isotope is replaced by the spin-1/2 $^{13}$C isotope, which couples to other nonzero-spin nuclei such as $^1$H (spin 1/2). Our particular focus is on the indirect couplings mediated by electrons shared in chemical bonds, known as $J$-couplings.
The presence or absence and relative amplitudes of peaks in ZULF spectra will depend on the choice of experimental geometry and protocol~\cite{Barskiy2025}.

In the following, we make several assumptions to simplify the discussion and mathematical description: (i) The spin-state preparation and detection geometry are assumed to be such that oscillating signals due to $J$-coupling are observable (an example is given in Fig.\,S2); (ii) the case of thermal polarization is considered, where spins are prepolarized by a permanent magnet or electromagnet. The prepolarizing field is then either switched off (in situ measurement~\cite{Storm2017,Bodenstedt2021}), or the spin sample is mechanically shuttled to an ultralow-field environment for evolution and detection (as in this work). Such shuttling may be either sudden or adiabatic. In the latter case, an appropriate pulse must be applied to produce a detectable signal (see the Supporting Information, SI, for further details)---here we assume pulse selection such that positive and negative $J$-peaks have equal amplitude at zero field. Consideration of hyperpolarization~\cite{Kovtunov2018,Eills2023} is beyond the scope of this work, as more complicated spin states are created in that case; however, extension of the theoretical description is possible. (iii) Detection of nZF peaks is omitted here, as our focus is on $J$-spectroscopy---a simplified detection operator is used where only the difference of spin operators is considered---but this topic is treated in the SI. 

Following Eqs.\,(18)--(19) in the SI, 
the expected signal of $J$-peaks can be written as
\begin{equation}
S_{x} = 
\cos \left(J t\right) \cos \left(\frac{\Sigma}{2} t\right) = 
\frac{1}{4} \left[ 
e^{i (J+\frac{\Sigma}{2}) t } + 
e^{-i (J+\frac{\Sigma}{2}) t } + 
e^{i (J-\frac{\Sigma}{2}) t } + 
e^{-i (J-\frac{\Sigma}{2}) t } \right] \,.
\end{equation}
Note that $J$ is defined in units of rad/s (see SI), and $\Sigma = \left(\gamma_{\rm H} + \gamma_{\rm C}\right) B_{z}$, where the gyromagnetic ratios 
are $\gamma_{\text{H}}/{2\pi} = 42.6$~Hz/$\upmu$T and $\gamma_{\text{C}}/{2\pi} = 10.7$~Hz/$\upmu$T. The latter term is responsible for the splitting of the \textit{J}-lines and also produces nZF lines, as described previously~\cite{Ledbetter2011} and fully detailed in the SI. 
Ignoring the nZF lines and without quadrature detection, the $x$-axis measurement contains all four $J$ frequency components (top spectrum in Fig.\,\ref{fig:quad}c). The application of laboratory-frame quadrature detection~\cite{Fabricant2024} allows for the separation of $J$-signals into their counter-rotating components. If the perpendicular ($y$-axis) signal is recorded simultaneously as
\begin{equation} 
S_y = \cos \left(J t\right) \sin \left(\frac{\Sigma}{2} t\right) = \frac{1}{4i} \left[ e^{i (J+\frac{\Sigma}{2}) t } - e^{-i (J+\frac{\Sigma}{2}) t } - e^{i (J-\frac{\Sigma}{2}) t } + e^{-i (J-\frac{\Sigma}{2}) t } \right] \,,
\end{equation}
the full quadrature signal can be constructed as
\begin{equation}
S_x + i S_y = 
\cos \left(J t\right) \left[ \cos \left(\frac{\Sigma}{2} t\right) + i \cdot \sin \left(\frac{\Sigma}{2} t\right) \right] 
= \frac{1}{2} 
\left[ e^{i (J+\frac{\Sigma}{2}) t } + 
e^{- i (J - \frac{\Sigma}{2}) t } \right] \,.
\end{equation}
This quadrature signal contains only two rotating frequencies (one positively rotating and one negatively rotating), greatly simplifying spectral analysis by
removing the symmetry around zero frequency, thus decrowding complicated ZULF-NMR spectra (Fig.\,\ref{fig:quad}c) while providing absolute frequency information.

To validate theoretical predictions, experimental data were collected using a setup and protocol already described in detail in prior work~\cite{Fabricant2024,Fabricant2026} and depicted schematically in Fig.\,S2; the detection scheme is shown in Fig.\,\ref{fig:quad}a.
Each sample was first prepolarized in a 1\,T magnet for 10\,s and then mechanically shuttled, within 100\,ms, by 36\,cm to reach the measurement region. Subsequently, a 30\,$\upmu$T DC magnetic-field pulse was applied along the $y$-axis, with a duration selected to initialize the desired spin state. Finally, spin evolution in the $x$-$y$ plane was detected for up to 10\,s with the quadrature detectors (Fig.\,\ref{fig:quad}). This $\sim$20\,s duty cycle was repeated many times for signal averaging, in order to reduce noise contributions (such as 50\,Hz power-line noise and harmonics thereof). In all experiments, a magnetic field of approximately 230\,nT was applied along the $z$-axis inside the piercing solenoid, corresponding to a 10\,Hz proton precession frequency during measurement.

The ability of quadrature detection to distinguish positive and negative frequency components of a signal can be understood by following the basic processing protocol for data collected from our two dual-axis detectors, which provide four different signals in total: $S_{x_1}$ and $S_{y_1}$ come from sensor 1 while $S_{x_2}$ and $S_{y_2}$ come from the sensor 2 (Fig.\,\ref{fig:quad}a). Let us consider the simplest case of a signal oscillation at one angular frequency $\omega_0$---for example, due to Larmor precession of a nuclear spin in the measurement field, or the Zeeman transitions of a typical ZULF-NMR sample containing small organic molecules. 
The precession frequency is conventionally defined to be $\omega_0 = -\gamma\,B_z$, where the minus sign arises from the conventional directions of magnetic moment (right-hand rule) and current.
Referring to Fig.\,\ref{fig:quad} and assuming that magnetization is initialized along $-\hat{x}$, the time-dependent magnetic fields measured at the four magnetometer channels are
\begin{equation}
\label{eq:single}
\begin{aligned}
S_{x_1}(t) &= \frac{1}{2}B\cos(\omega_0 t)e^{-t/T_2}\,,  &\qquad
S_{y_1}(t) &= -B\sin(\omega_0 t)e^{-t/T_2}\,, \\
S_{x_2}(t) &= -B\cos(\omega_0 t)e^{-t/T_2}\,,           &\qquad
S_{y_2}(t) &= \frac{1}{2}B\sin(\omega_0 t)e^{-t/T_2}\,. 
\end{aligned} 
\end{equation} 
Here we have assumed that the sample magnetization may be well-approximated as a magnetic dipole~\cite{Fabricant2024,Fabricant2026}; $B$ is the maximum measured field, $t$ is the elapsed measurement time, and $T_2$ is the coherence time of the signal. Equation~\eqref{eq:single} may be obtained from a single measurement or the average of many measurements. 
The factors of 1/2 present for $S_{x_1}$  and $S_{y_2}$ stem from the fact that the sensors are oriented such that their longitudinal axes are radial to the sample.
Accordingly, due to the nature of dipolar fields, the magnetic response of sensor 1 and sensor 2 to the $x$- and $y$-fields, respectively, is only half that of the other sensitive channel (Fig.\,\ref{fig:quad}a). 

After signal acquisition, the gradiometric channels $S_{x}$ and $S_{y}$ are constructed as the difference of the individual channels:
\begin{align}
    S_x\left(t\right) = S_{x_1}\left(t\right) - S_{x_2}\left(t\right) \,, \qquad
    S_y\left(t\right) = S_{y_1}\left(t\right) - S_{y_2}\left(t\right) \,. \label{eq:grad}
\end{align}
Because the sine function is odd in $\omega_0$, $S_{y_1}$, $S_{y_2}$, and $S_{y}$ experience a sign change under inversion of the frequency coordinate.
The goal of this gradiometric subtraction without relative scaling is to remove common-mode systematic noise which does not come from the sample.
Finally, the quadrature signal is constructed from $S_{x}$ and $S_{y}$ as
\begin{equation}
    S\left(t\right) = e^{i\phi} \left[ S_x\left(t\right) - i\,S_y\left(t\right) \right]
    = 
    1.5 B\,e^{i\left(\omega_0 t + \phi\right) - t/T_2} \,,
    \label{eq:quad} 
\end{equation}
where $\phi$ is an overall phase. The factor $ 1.5 B $ results from the summation of the individual sensors contributing $ B $ and $ 0.5 B $, respectively.
We also see that the direction of rotation of ${S}\left(t\right)$ in the complex $x$-$y$ plane depends on the sign of $\omega_0$. The sign (positive or negative) used in construction of the quadrature signal from gradiometric channels corresponds to the sign of the detection field $B_z$ (Fig.\,S2).

\begin{figure}
  \includegraphics[width=\linewidth]{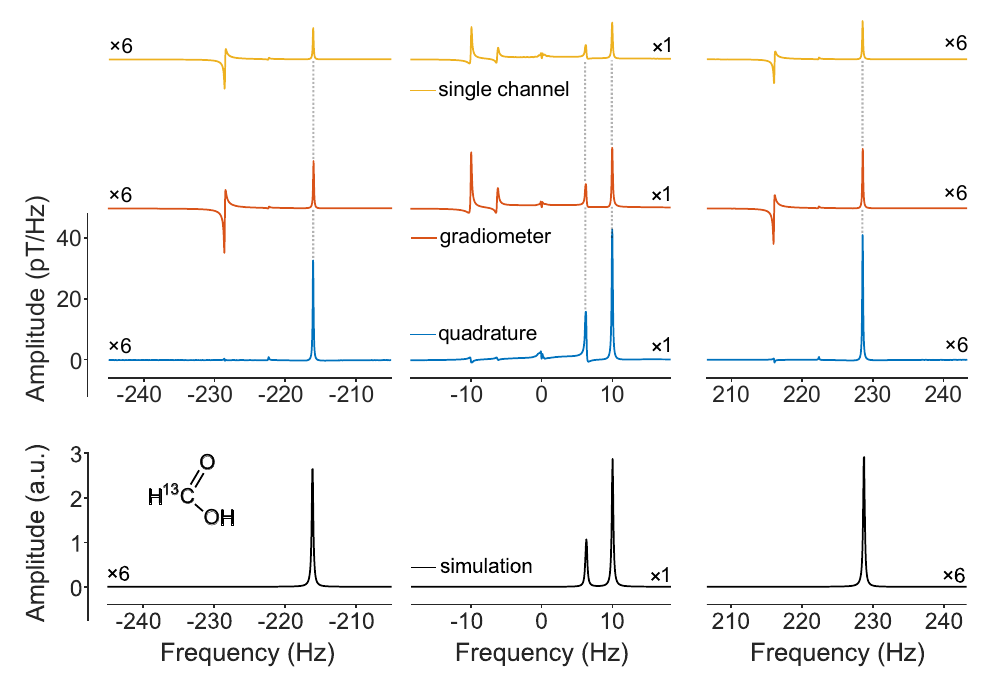}
  \caption{Measured and simulated ZULF-NMR spectra of an aqueous $\sim$25.1\,M solution of $^{13}$C-labeled formic acid (Sigma-Aldrich 279404) at a background field of $B_z=230$\,nT. 
  The energy-level diagram for this XA spin system is depicted in Fig.\,\ref{fig:quad}.
  Experimental spectra (top three rows) were obtained from the average of 4096 scans. Single-channel ($y_1$), gradiometric ($y$), and quadrature traces are displayed in yellow, red, and blue, respectively; as a visual aid, vertical dashed gray lines connect corresponding signals. The duration of the 30\,$\upmu$T $y$-pulse (Fig.\,S2) was optimized to be 200\,$\upmu$s for the nZF spectrum (central panel) and 550\,$\upmu$s for the $J$-spectra (outermost panels). Each spectral region was phased individually using the following relative zero-order phase corrections: from left to right, $\{\frac{3\pi}{10},\,\frac{9\pi}{10},\,-\frac{19\pi}{20}\}$ (single channel and gradiometric data) and $\{\frac{\pi}{5},\,-\frac{2\pi}{5},\,-\frac{11\pi}{20}\}$ (quadrature data). For constructing the quadrature data, the relative gains of the two gradiometric channels were adjusted by 5\% to suppress residual peaks due to differing sensor gains ~\cite{Fabricant2024,Elzenheimer2025} (see SI). 
  The small feature at $\pm J_{\rm Hz}$, where $J_{\rm Hz}=222.4$\,Hz, arises from stray $x$-$y$ fields; artifacts around 0\,Hz are attributed to non-common-mode noise not canceled by gradiometry. Prior to application of lineshape broadening to all spectra for SNR enhancement, the linewidth of the $J$-peaks around $\pm230$\,Hz was found to be $\sim100$\,mHz, with an SNR of over 3000.
  Quadrature simulation (bottom row) includes dipolar relaxation and other realistic experimental parameters, including the $z$-field profile during shuttling and $y$-pulse application in the presence of the $B_z$ field.} \label{fig:formicacid}
\end{figure}

\begin{figure}
  \includegraphics[trim=0pt 30pt 00pt 0pt, clip, width=\linewidth]{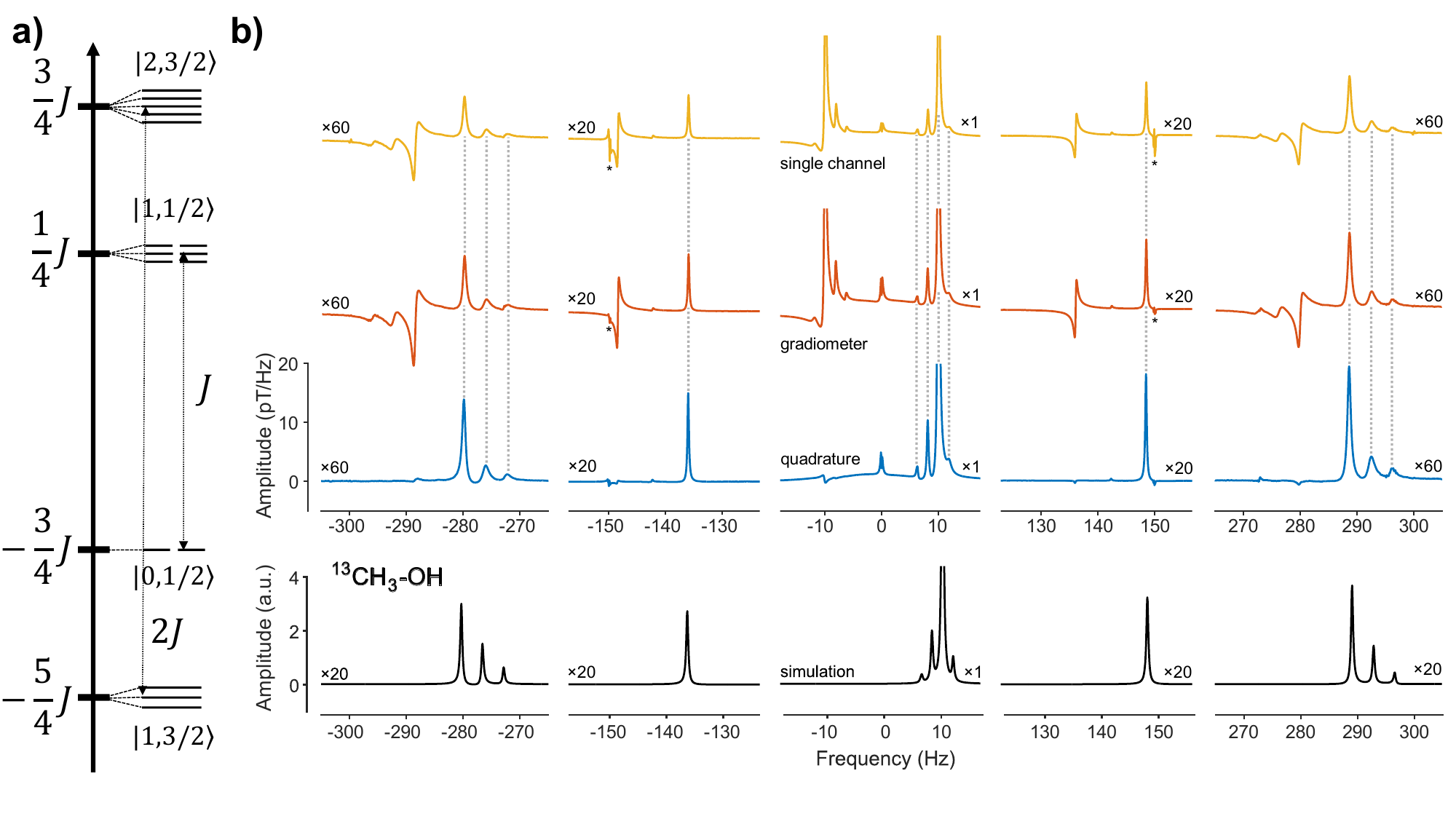}
  \caption{
  a) Energy-level diagram of the $^{13}$CH$_3$ spin system at ultralow field showing overpopulated (filled circles) and underpopulated (empty circles) states after adiabatic transport of high-field polarization. 
  See Fig.\,S3 for a more detailed energy-level diagram relating frequencies to spectral lines. b) Measured and simulated ZULF-NMR spectra of an aqueous solution of $^{13}$C-labeled methanol (8\,M $^{13}$CH$_3$OH in H$_2$O, 1.5 mL) at a background field of {$B_z=230$\,nT}.
  Experimental spectra (top three rows) were obtained from the average of 4096 scans. Single-channel ($y_1$), gradiometric ($y$), and quadrature traces are displayed in yellow, red, and blue, respectively; as a visual aid, vertical dashed gray lines connect corresponding signals. The duration of the 30\,$\upmu$T $y$-pulse was optimized to be 150\,$\upmu$s for the nZF spectrum (central panel) and 550\,$\upmu$s for the $J$- and $2J$-spectra (outermost panels). Each spectral region was phased individually using the following relative zero-order phase corrections: from left to right, $\{\frac{2\pi}{5},\,\frac{\pi}{4},\,-\frac{9\pi}{10},\,\frac{\pi}{4},\,\frac{3\pi}{10}\}$ (single channel and gradiometer) and $\{\frac{4\pi}{5},\,\frac{3\pi}{4},\,-\frac{2\pi}{5},\,\frac{3\pi}{4},\,\frac{4\pi}{5}\}$ (quadrature). For constructing the quadrature data, the relative gains of the two gradiometric channels were adjusted by 7\% to suppress residual peaks due to differing sensor gains ~\cite{Fabricant2024,Elzenheimer2025} (see SI). The small features at $\pm J_{\rm Hz}$  and $\pm2J_{\rm Hz}$, where $J_{\rm Hz}=142.3$\,Hz arise from stray $x$-$y$ fields; artifacts around 0\,Hz are attributed to non-common-mode sensor noise not canceled by gradiometry. Prior to application of lineshape broadening to all spectra for SNR enhancement, the linewidths of the $J$-peaks around $\pm140$\,Hz and the $2J$-peaks around $\pm280$\,Hz were found to be $\sim$200\,mHz and $\sim$600\,mHz, respectively.
  Quadrature simulation (bottom row) includes dipolar relaxation and other realistic experimental parameters, including the $z$-field profile during shuttling and $y$-pulse application in the presence of the $B_z$ field.
  }
  \label{fig:methanol}
\end{figure}

Taking the Fourier transform (FT) of Eq.\,\eqref{eq:quad}, we find
\begin{equation}
    S\left(\omega\right) = \int_0^{\infty} S\left(t\right)\,e^{-i\omega t}\,dt \,=
    \frac{1.5 B \, e^{i\phi}}{i\left(\omega-\omega_0\right) + 1/T_2} \,,
    \label{eq:Fourier}
\end{equation}
which we recognize as a complex Lorentzian function centered on $\omega=\omega_0$, where $\omega_0$ can be positive or negative. 
If a single-channel or gradiometric signal is used in Eq.\,\eqref{eq:Fourier} instead of a quadrature signal, the integral will contain a term proportional to $\left[ e^{-i\left(\omega-\omega_0\right)t} \pm e^{-i\left(\omega+\omega_0\right)t} \right]/2$ rather than $e^{-i\left(\omega-\omega_0\right)t}$, such that the FT is symmetric about zero and information about the sign of $\omega_0$ is not available. Furthermore, we see that gradiometric quadrature detection enhances the signal amplitude by a factor of 3 over single-axis measurement (the $x_2$ and $y_1$ channels in Eq.\,\eqref{eq:single}).
This factor of 3 is composed of a factor of 3/2 from construction of the gradiometric signals (Eq.\,\eqref{eq:grad}) and a factor of 2 from construction of the quadrature signal (Eq.\,\eqref{eq:quad}), which concentrates the signal at its actual frequency instead of distributing intensity into positive and negative components. 
As discussed above and demonstrated in Figs.\,\ref{fig:formicacid}--\ref{fig:methanol}, signals may consist of multiple positive and negative frequencies contributions, depending on the prepared spin state and nutation angle induced by the detection pulse.\cite{Teleanu2025}


Note that in actual processing of our ZULF-NMR data, recorded by digitizing analog sensor outputs, the relative gains of the gradiometric channels in Eq.\,\eqref{eq:quad} may be adjusted in order to account for sensor gain variability and imperfections of the experimental geometry~\cite{Fabricant2024,Osborne2018}, such that the negative-frequency residuals of nZF peaks (for systems with $\gamma>0$) are suppressed as much as possible. In order to minimize misalignment of the sensors, they are contained in a single 3D-printed structure. However, variability in gain and frequency response of commercially available atomic magnetometers~\cite{Elzenheimer2025} typically necessitates gain adjustments of 0--10\% in construction of the quadrature signal.


To experimentally validate the phase-sensitive nature of quadrature detection in ZULF NMR, we begin with the simplest case of an XA (two-spin) system, [$^{13}$C]-formic acid. This molecule features a heteronuclear spin-1/2 pair with a $J$-coupling of 222\,Hz \cite{Theis2013,Emondts2014,Put2021}. As shown in Fig.\,\ref{fig:formicacid}, a sample of the [$^{13}$C]-formic acid produces two groups of peaks in the spectra when a magnetic field of 230\,nT is applied perpendicular to the sensing axes. The first group (nZF peaks) corresponds to transitions within the triplet manifold of the coupled $^1$H-$^{13}$C pair at 6.3\,Hz (Fig.\,\ref{fig:formicacid}), and a peak from the uncoupled protons precessing at 10.0\,Hz. The second group contains the higher-frequency $J$-peaks corresponding to transitions between the singlet and triplet states (Fig.\,\ref{fig:quad}b). Matching simulations shown in Fig.\,\ref{fig:formicacid} were obtained using the Spinach package in MATLAB with realistic experimental parameters (see SI).

Importantly, the detectability of these peaks depends on the symmetry of the initial spin state. The nZF peaks originate from initial magnetization terms proportional to $(\hat{I}_{1x} + \hat{I}_{2x})$, whereas the $J$ peaks stem from antisymmetric terms proportional to $(\hat{I}_{1x} - \hat{I}_{2x})$. When using single-axis detection, all peaks (nZF and $J$-peaks) are visible at negative and positive frequencies. However, the application of quadrature detection refines the spectrum. The nZF peaks will only appear on one side of the spectrum (depending on $\gamma$), and the $J$-peaks are split in two: one positive-frequency peak at $J + \Sigma/2$ and one negative-frequency peak at $-J + \Sigma/2$ (Figs.\,\ref{fig:quad}c and \ref{fig:formicacid}), consistent with the simple theoretical model presented here as well as with the more mathematically detailed density-matrix analysis provided in the SI.

It should be noted that single-channel ZULF spectra often suffer from phase distortions that are not resolved by gradiometry alone~\cite{Jiang2019}. For example, the different nutation patterns of individual peaks in our experimental modality (see Figs.\,S4--S5) result in different phases of $J$-peaks in the single-channel and gradiometric spectra of Fig.\,\ref{fig:formicacid}. Thus, when the positive-frequency (or negative-frequency) spectrum is phased to obtain an absorptive lineshape of the rightmost $J$-peak, the corresponding leftmost peak appears out of phase (dispersive). This artifact is eliminated by applying quadrature processing due to the separation of the $+J$ and $-J$ manifolds into positive and negative frequency space, illustrating yet another benefit of truly phase-sensitive detection in the laboratory frame. 
Additional practical advantages of our QD method include the simplification of magnetic-field shimming procedures at zero field. Stray DC fields must be removed in a complicated procedure in order to avoid unwanted splitting of overlapping $J$-peaks when collecting non-quadrature zero-field spectra,~\cite{Sjolander2017} but this is not necessary in the quadrature case where the $\pm J$-peaks are already resolved by their opposite signs in frequency space. 

To further demonstrate the utility of quadrature detection even in more complex spin systems, we analyzed [$^{13}$C]-methanol as a representative XA$_3$-type spin system. The experimental and simulated spectra (Fig.\,\ref{fig:methanol}) display three families of peaks: nZF peaks and two sets of $J$-peaks centered around the $J$ and $2J$ frequencies, as expected from the underlying spin topology~\cite{Butler2013,Theis2013,Myers2024}. Unlike the simpler case of the XA spin system, here the nZF peaks now display overlap of signals from uncoupled water or hydroxyl protons.
Overall, the methanol data provided in (Fig.\,\ref{fig:methanol}) display properties similar to those observed for formic acid. Namely, the nZF peaks can be observed as purely positive-frequency peaks in the quadrature spectrum, because the gyromagnetic ratios of both nuclear species are positive. The doublet centered around $J$ divides into positive and negative peaks in the quadrature spectrum, such that the peak at $J+\frac{\Sigma}{2}$ becomes positive, and the peak at $J-\frac{\Sigma}{2}$ becomes negative.
For the peaks centered around $2J$, additional behavior is observed. In the single-channel spectrum, a sextet is measured, with the pair closest to the center given by $2J \pm B_{\perp} \cdot \left(\frac{1}{4} \gamma_{\rm H} + \frac{3}{4} \gamma_{\rm C}\right)$ in analogy to the splitting of the $J$-lines; the four other lines become visible at frequencies given by
$2J \pm B_{\perp} \cdot \left(\frac{3}{4} \gamma_{\rm H} + \frac{1}{4} \gamma_{\rm C}\right)$
and
$2J \pm B_{\perp} \cdot \left(\frac{5}{4}\gamma_{\rm H} - \frac{1}{4}\gamma_{\rm C}\right)$.\cite{Ledbetter2011,Appelt2010}. 
In the quadrature spectrum, this $2J$ sextet divides into two groups, where the three peaks having an absolute frequency $>2J$ now appear only at positive frequencies and those having an absolute frequency $<2J$ appear only at negative frequencies, as seen in Fig.\,\ref{fig:methanol}. As before, the simulations match the experiments, with peaks appearing at 288.6\,Hz, 292.5\,Hz, and 296.2\,Hz in the experimental $+2J$-spectrum. The observed XA$_3$ spin-system dynamics in [$^{13}$C]-methanol provide another example of how quadrature ZULF-NMR spectroscopy may avoid spectral overcrowding---especially when considering more complex systems or chemical mixtures in the future, such as those presented in a recent \textit{in silico} ZULF-NMR study of more than 200 molecules~\cite{Mandzhieva2025}.

To conclude, in this work we have demonstrated the concept of quadrature-detected zero-to-ultralow-field (QD-ZULF) NMR.
By recording orthogonal components of oscillating nuclear magnetization in the laboratory frame and recombining them into a complex signal, one can recover absolute phase information, thus distinguishing the signs of gyromagnetic ratios and other frequency-dependent information. 
Experiments on thermally polarized [$^{13}$C]-formic acid and [$^{13}$C]-methanol validate theory and show excellent agreement with simulations, demonstrating frequency separation of counter-rotating $J$-components as well as simplification of crowded multiplets. 
Our phase-sensitive, gradiometric quadrature approach can streamline spin-system assignments and paves the way toward hyperpolarized and multidimensional zero-field methods. 
Implementation of QD ZULF NMR will facilitate richer understanding of magnetization dynamics, including through the analysis of nutation behavior of spin systems under variable pulse conditions~\cite{Salikhov2022} as well as through studies of relaxation mechanisms of different resonance lines~\cite{Teleanu2025}. 


\clearpage
\begin{acknowledgement}


The authors thank Peter Blümler for providing the polarizing magnet and Sergio Iglesias Noval for collecting calibration spectra, and acknowledge helpful discussions with John Blanchard, James Eills, Chengtong Zhang, Igor Koptyug, and Rom\'{a}n Picazo-Frutos. This project was supported in part by the Deutsche Forschungsgemeinschaft (DFG, German Research Foundation) under grant BU 3035/15-1, and in the framework of the collaborative research center “Defects and Defect Engineering in Soft Matter” (SFB1552) under Project No. 465145163. 
F.T. acknowledges funding from National Medical Project MySMIS 326475 and Project ELI-RO/RDI/2024/14 SPARC funded by the Institute of Atomic Physics (Romania). A.J. acknowledges funding from the US National Science Foundation, award no. CHE 2505792, an award from the ACS, PRF 68117-ND6, the Carl Zeiss Humboldt Research Award, and partially by an unrestricted gift from Google.
D.A.B. acknowledges support from the Alexander von Humboldt Foundation in the framework of the Sofja Kovalevskaja Award, and the University of Miami Provost's research award.

\end{acknowledgement}

\clearpage
\begin{suppinfo}


The following is available free of charge: figures showing distinguishability of counter-rotating NMR signals, experimental apparatus and protocol, theory comparison of the XA spin system in different field regimes, energy-level diagram of the XA$_3$ spin system, and experimental nutation curves; in-depth discussion of the frequencies of near-zero-field NMR signals for a heteronuclear two-spin system; data used to produce the spectra in Figs.\,\ref{fig:formicacid}--\ref{fig:methanol}, as well as the corresponding data-analysis and simulation notebooks.

\end{suppinfo}

\clearpage
\bibliography{quadrature}

@article{Eills2023,
   author = {James Eills and Dmitry Budker and Silvia Cavagnero and Eduard Y. Chekmenev and Stuart J. Elliott and Sami Jannin and Anne Lesage and Jörg Matysik and Thomas Meersmann and Thomas Prisner and Jeffrey A. Reimer and Hanming Yang and Igor V. Koptyug},
   doi = {10.1021/acs.chemrev.2c00534},
   issn = {0009-2665},
   issue = {4},
   journal = {Chemical Reviews},
   month = {2},
   pages = {1417-1551},
   pmid = {36701528},
   publisher = {American Chemical Society},
   title = {Spin Hyperpolarization in Modern Magnetic Resonance},
   volume = {123},
   url = {https://pubs.acs.org/doi/10.1021/acs.chemrev.2c00534},
   year = {2023}
}

@article{Kovtunov2018,
   author = {Kirill V. Kovtunov and Ekaterina V. Pokochueva and Oleg G. Salnikov and Samuel F. Cousin and Dennis Kurzbach and Basile Vuichoud and Sami Jannin and Eduard Y. Chekmenev and Boyd M. Goodson and Danila A. Barskiy and Igor V. Koptyug},
   doi = {10.1002/asia.201800551},
   issn = {1861-4728},
   issue = {15},
   journal = {Chemistry – An Asian Journal},
   month = {8},
   pages = {1857-1871},
   publisher = {John Wiley \& Sons, Ltd},
   title = {Hyperpolarized NMR Spectroscopy: <i>d</i> ‐DNP, PHIP, and SABRE Techniques},
   volume = {13},
   url = {https://onlinelibrary.wiley.com/doi/10.1002/asia.201800551},
   year = {2018}
}

@article{Appelt2010,
   author = {S. Appelt and F. W. Häsing and U. Sieling and A. Gordji-Nejad and S. Glöggler and B. Blümich},
   doi = {10.1103/PhysRevA.81.023420},
   issn = {1050-2947},
   issue = {2},
   journal = {Physical Review A},
   month = {2},
   pages = {023420},
   publisher = {American Physical Society},
   title = {Paths from weak to strong coupling in NMR},
   volume = {81},
   url = {https://link.aps.org/doi/10.1103/PhysRevA.81.023420},
   year = {2010}
}

@article{Bodenstedt2021,
   author = {Sven Bodenstedt and Morgan W. Mitchell and Michael C. D. Tayler},
   doi = {10.1038/s41467-021-24248-9},
   issn = {2041-1723},
   issue = {1},
   journal = {Nature Communications},
   month = {6},
   pages = {4041},
   pmid = {34193862},
   publisher = {Nature Publishing Group},
   title = {Fast-field-cycling ultralow-field nuclear magnetic relaxation dispersion},
   volume = {12},
   url = {https://www.nature.com/articles/s41467-021-24248-9},
   year = {2021}
}

@article{Blanchard2021,
   author = {John W. Blanchard and Dmitry Budker and Andreas Trabesinger},
   doi = {10.1016/j.jmr.2020.106886},
   issn = {10907807},
   journal = {Journal of Magnetic Resonance},
   month = {2},
   pages = {106886},
   pmid = {33518173},
   publisher = {Academic Press},
   title = {Lower than low: Perspectives on zero- to ultralow-field nuclear magnetic resonance},
   volume = {323},
   url = {https://linkinghub.elsevier.com/retrieve/pii/S1090780720302044},
   year = {2021}
}

@inbook{BlanchardBudker2016,
   author = {John W. Blanchard and Dmitry Budker},
   city = {Chichester, UK},
   doi = {10.1002/9780470034590.emrstm1369},
   issn = {20556101},
   issue = {3},
   booktitle = {eMagRes},
   month = {9},
   pages = {1395-1410},
   publisher = {John Wiley \& Sons, Ltd},
   title = {Zero- to Ultralow-Field NMR},
   volume = {5},
   url = {http://doi.wiley.com/10.1002/9780470034590.emrstm1369},
   year = {2016}
}

@article{Osborne2018,
   author = {James Osborne and Jeff Orton and Orang Alem and Vishal Shah},
   doi = {10.1117/12.2299197},
   editor = {Selim M. Shahriar and Jacob Scheuer},
   isbn = {9781510615816},
   issn = {1996756X},
   journal = {Proc. SPIE 10548, Steep Dispersion Engineering and Opto-Atomic Precision Metrology XI},
   month = {2},
   pages = {51},
   publisher = {SPIE},
   title = {Fully integrated, standalone zero field optically pumped magnetometer for biomagnetism},
   volume = {10548G},
   url = {https://www.spiedigitallibrary.org/conference-proceedings-of-spie/10548/2299197/Fully-integrated-standalone-zero-field-optically-pumped-magnetometer-for-biomagnetism/10.1117/12.2299197.full},
   year = {2018}
}

@book{Keeler2010,
   author = {James Keeler},
   edition = {2},
   isbn = {978-0-470-74608-0},
   publisher = {John Wiley \& Sons, Ltd},
   title = {Understanding NMR Spectroscopy},
   url = {https://www.wiley.com/en-us/Understanding+NMR+Spectroscopy%2C+2nd+Edition-p-9780470746080},
   year = {2010}
}

@book{Levitt2008,
   author = {Malcolm H. Levitt},
   edition = {2},
   isbn = {978-0-470-51117-6},
   issn = {1098-6596},
   journal = {Wiley},
   month = {3},
   pmid = {25246403},
   publisher = {John Wiley \& Sons, Ltd},
   title = {Spin Dynamics: Basics of Nuclear Magnetic Resonance},
   url = {https://www.wiley.com/en-ae/Spin+Dynamics%3A+Basics+of+Nuclear+Magnetic+Resonance%2C+2nd+Edition-p-9780470511176},
   year = {2008}
}

@article{Salikhov2022,
   author = {K. M. Salikhov},
   doi = {10.1134/S1063776122110164},
   issn = {1063-7761},
   issue = {5},
   journal = {Journal of Experimental and Theoretical Physics},
   month = {11},
   pages = {617-631},
   publisher = {Pleiades Publishing},
   title = {A New View on Spin “Nutation”},
   volume = {135},
   url = {https://link.springer.com/10.1134/S1063776122110164},
   year = {2022}
}

@article{Fabricant2024,
   author = {Anne M. Fabricant and Piotr Put and Danila A. Barskiy},
   doi = {10.3389/fpls.2024.1352282},
   issn = {1664-462X},
   journal = {Frontiers in Plant Science},
   month = {3},
   pages = {1352282},
   publisher = {Frontiers Media SA},
   title = {Proton relaxometry of tree leaves at hypogeomagnetic fields},
   volume = {15},
   url = {https://www.frontiersin.org/articles/10.3389/fpls.2024.1352282/full},
   year = {2024}
}

@article{Levitt1997,
   author = {Malcolm H. Levitt},
   doi = {10.1006/jmre.1997.1161},
   issn = {10907807},
   issue = {2},
   journal = {Journal of Magnetic Resonance},
   month = {6},
   pages = {164-182},
   publisher = {Academic Press},
   title = {The Signs of Frequencies and Phases in NMR},
   volume = {126},
   url = {https://linkinghub.elsevier.com/retrieve/pii/S1090780797911617},
   year = {1997}
}

@article{Storm2017,
   author = {Jan-Hendrik Storm and Peter Hömmen and Dietmar Drung and Rainer Körber},
   doi = {10.1063/1.4976823},
   issn = {0003-6951},
   issue = {7},
   journal = {Applied Physics Letters},
   month = {2},
   pages = {72603},
   publisher = {American Institute of Physics Inc.},
   title = {An ultra-sensitive and wideband magnetometer based on a superconducting quantum interference device},
   volume = {110},
   url = {https://pubs.aip.org/apl/article/110/7/072603/33797/An-ultra-sensitive-and-wideband-magnetometer-based},
   year = {2017}
}

@article{Teleanu2025,
   author = {Florin Teleanu and Anne M. Fabricant and Chengtong Zhang and Gary P. Centers and Dmitry Budker and Danila A. Barskiy and Alexej Jerschow},
   month = {11},
   title = {Nuclear spin relaxation in zero- to ultralow-field magnetic resonance spectroscopy},
   url = {http://arxiv.org/abs/2511.08517},
   year = {2025}
}

@article{Ledbetter2011,
   author = {M. P. Ledbetter and T. Theis and J. W. Blanchard and H. Ring and P. Ganssle and S. Appelt and B. Blümich and A. Pines and D. Budker},
   doi = {10.1103/PhysRevLett.107.107601},
   issn = {0031-9007},
   issue = {10},
   journal = {Physical Review Letters},
   month = {9},
   pages = {107601},
   pmid = {21981529},
   publisher = {American Physical Society},
   title = {Near-Zero-Field Nuclear Magnetic Resonance},
   volume = {107},
   url = {https://link.aps.org/doi/10.1103/PhysRevLett.107.107601},
   year = {2011}
}

@article{Theis2013,
   author = {Thomas Theis and John W. Blanchard and Mark C. Butler and Micah P. Ledbetter and Dmitry Budker and Alexander Pines},
   doi = {10.1016/j.cplett.2013.06.042},
   issn = {00092614},
   journal = {Chemical Physics Letters},
   month = {8},
   pages = {160-165},
   publisher = {North-Holland},
   title = {Chemical analysis using J-coupling multiplets in zero-field NMR},
   volume = {580},
   url = {https://linkinghub.elsevier.com/retrieve/pii/S0009261413008191},
   year = {2013}
}

@article{Doty1999,
   author = {F.David Doty and George Entzminger and Cory D. Hauck and John P. Staab},
   doi = {10.1006/jmre.1998.1703},
   issn = {10907807},
   issue = {1},
   journal = {Journal of Magnetic Resonance},
   month = {5},
   pages = {144-154},
   pmid = {10329237},
   publisher = {Academic Press},
   title = {Practical Aspects of Birdcage Coils},
   volume = {138},
   url = {https://linkinghub.elsevier.com/retrieve/pii/S1090780798917037},
   year = {1999}
}

@article{King2017,
   author = {Jonathan P. King and Tobias F. Sjolander and John W. Blanchard},
   doi = {10.1021/acs.jpclett.6b02653},
   issn = {1948-7185},
   issue = {4},
   journal = {The Journal of Physical Chemistry Letters},
   month = {2},
   pages = {710-714},
   pmid = {28029791},
   publisher = {American Chemical Society},
   title = {Antisymmetric Couplings Enable Direct Observation of Chirality in Nuclear Magnetic Resonance Spectroscopy},
   volume = {8},
   url = {https://pubs.acs.org/doi/10.1021/acs.jpclett.6b02653},
   year = {2017}
}

@article{Georgiou2024,
   author = {T. Georgiou and J. L. Palma and V. Mujica and S. Varela and M. Galante and V. J. Santamaría-García and L. Mboning and R. N. Schwartz and G. Cuniberti and L.-S. Bouchard},
   doi = {10.1038/s41467-024-49966-8},
   issn = {2041-1723},
   issue = {1},
   journal = {Nature Communications},
   month = {8},
   pages = {7367},
   pmid = {39191753},
   publisher = {Nature Publishing Group},
   title = {Enantiospecificity in NMR enabled by chirality-induced spin selectivity},
   volume = {15},
   url = {https://www.nature.com/articles/s41467-024-49966-8},
   year = {2024}
}

@article{Wadhwa2024,
   author = {Sagar Wadhwa and Dominique Buyens and Jan G. Korvink},
   doi = {10.1002/adma.202408547},
   issn = {0935-9648},
   issue = {40},
   journal = {Advanced Materials},
   month = {10},
   pages = {2408547},
   pmid = {39180269},
   publisher = {John Wiley and Sons Inc},
   title = {Direct Chiral Discrimination with NMR},
   volume = {36},
   url = {https://advanced.onlinelibrary.wiley.com/doi/10.1002/adma.202408547},
   year = {2024}
}

@article{Barskiy2025,
   author = {Danila A. Barskiy and John W. Blanchard and Dmitry Budker and James Eills and Szymon Pustelny and Kirill F. Sheberstov and Michael C.D. Tayler and Andreas H. Trabesinger},
   doi = {10.1016/j.pnmrs.2025.101558},
   issn = {00796565},
   journal = {Progress in Nuclear Magnetic Resonance Spectroscopy},
   month = {8},
   pages = {101558},
   pmid = {40912877},
   publisher = {Pergamon},
   title = {Zero- to ultralow-field nuclear magnetic resonance},
   volume = {148-149},
   url = {https://linkinghub.elsevier.com/retrieve/pii/S0079656525000020},
   year = {2025}
}

@article{Put2021,
   author = {Piotr Put and Szymon Pustelny and Dmitry Budker and Emanuel Druga and Tobias F. Sjolander and Alexander Pines and Danila A. Barskiy},
   doi = {10.1021/acs.analchem.0c04738},
   issn = {0003-2700},
   issue = {6},
   journal = {Analytical Chemistry},
   month = {2},
   pages = {3226-3232},
   publisher = {American Chemical Society},
   title = {Zero- to Ultralow-Field NMR Spectroscopy of Small Biomolecules},
   volume = {93},
   url = {https://pubs.acs.org/doi/10.1021/acs.analchem.0c04738},
   year = {2021}
}

@article{Emondts2014,
   author = {M. Emondts and M. P. Ledbetter and S. Pustelny and T. Theis and B. Patton and J. W. Blanchard and M. C. Butler and D. Budker and A. Pines},
   doi = {10.1103/PhysRevLett.112.077601},
   issn = {0031-9007},
   issue = {7},
   journal = {Physical Review Letters},
   month = {2},
   pages = {077601},
   publisher = {American Physical Society},
   title = {Long-Lived Heteronuclear Spin-Singlet States in Liquids at a Zero Magnetic field},
   volume = {112},
   url = {https://link.aps.org/doi/10.1103/PhysRevLett.112.077601},
   year = {2014}
}

@article{Ledbetter2009,
   author = {M.P. Ledbetter and C.W. Crawford and A. Pines and D.E. Wemmer and S. Knappe and J. Kitching and D. Budker},
   doi = {10.1016/j.jmr.2009.03.008},
   issn = {10907807},
   issue = {1},
   journal = {Journal of Magnetic Resonance},
   month = {7},
   pages = {25-29},
   pmid = {19406678},
   publisher = {Academic Press},
   title = {Optical detection of NMR J-spectra at zero magnetic field},
   volume = {199},
   url = {https://linkinghub.elsevier.com/retrieve/pii/S1090780709000780},
   year = {2009}
}

@article{Butler2013,
   author = {Mark C. Butler and Micah P. Ledbetter and Thomas Theis and John W. Blanchard and Dmitry Budker and Alexander Pines},
   doi = {10.1063/1.4803144},
   issn = {0021-9606},
   issue = {18},
   journal = {The Journal of Chemical Physics},
   month = {5},
   pages = {184202},
   publisher = {AIP Publishing},
   title = {Multiplets at zero magnetic field: The geometry of zero-field NMR},
   volume = {138},
   url = {https://pubs.aip.org/jcp/article/138/18/184202/985516/Multiplets-at-zero-magnetic-field-The-geometry-of},
   year = {2013}
}

@article{Jiang2019,
   author = {Min Jiang and Román Picazo Frutos and Teng Wu and John W. Blanchard and Xinhua Peng and Dmitry Budker},
   doi = {10.1103/PhysRevApplied.11.024005},
   issn = {2331-7019},
   issue = {2},
   journal = {Physical Review Applied},
   month = {2},
   pages = {024005},
   publisher = {American Physical Society},
   title = {Magnetic Gradiometer for the Detection of Zero- to Ultralow-Field Nuclear Magnetic Resonance},
   volume = {11},
   url = {https://link.aps.org/doi/10.1103/PhysRevApplied.11.024005},
   year = {2019}
}

@article{Mandzhieva2025,
   author = {Iuliia Mandzhieva and Franziska Theiss and Xingtao He and Adam Ortmeier and Anuja Koirala and Stephen J. McBride and Stephen J. DeVience and Matthew S. Rosen and Volker Blum and Thomas Theis},
   doi = {10.1021/acs.jcim.5c00111},
   issn = {1549-9596},
   issue = {14},
   journal = {Journal of Chemical Information and Modeling},
   month = {7},
   pages = {7554-7568},
   pmid = {40626657},
   publisher = {American Chemical Society},
   title = {Zero-Field NMR and Millitesla-SLIC Spectra for \&gt;200 Molecules from Density Functional Theory and Spin Dynamics},
   volume = {65},
   url = {https://pubs.acs.org/doi/10.1021/acs.jcim.5c00111},
   year = {2025}
}

@article{Elzenheimer2025,
   author = {Eric Elzenheimer and Silvia Knappe-Grüneberg and Jan Zerfowski and William Evans and Fabian Grüneberg and Michael Höft and Surjo R. Soekadar and Stephen Robinson and Jens Voigt},
   doi = {10.1109/JSEN.2024.3491164},
   issn = {1530-437X},
   issue = {2},
   journal = {IEEE Sensors Journal},
   month = {1},
   pages = {2432-2455},
   publisher = {Institute of Electrical and Electronics Engineers Inc.},
   title = {Key Metrics and Experimental Test Bench for Assessing Highly Sensitive Magnetometers in Research},
   volume = {25},
   url = {https://ieeexplore.ieee.org/document/10750198/},
   year = {2025}
}

@phdthesis{Sjolander2017,
   author = {Tobias F. Sjolander},
   school = {University of California, Berkeley},
   title = {Advances in Pulsed Zero-Field NMR},
   url = {https://escholarship.org/uc/item/2kj4v04n},
   year = {2017}
}

@article{Fabricant2026,
   author = {Anne M. Fabricant and Román Picazo-Frutos and Florin Teleanu and Gregory J. Rees and Raphael Kircher and Mengjiang Lin and William Evans and Paul-Martin Luc and Robert A. House and Peter G. Bruce and Peter Krüger and John W. Blanchard and James Eills and Kirill F. Sheberstov and Rainer Körber and Dmitry Budker and Danila A. Barskiy and Alexej Jerschow},
   doi = {10.1039/D5SC04419G},
   issn = {2041-6520},
   issue = {12},
   journal = {Chemical Science},
   month = {3},
   pages = {5877-5887},
   pmid = {41716241},
   publisher = {The Royal Society of Chemistry},
   title = {Enabling nondestructive observation of electrolyte composition in batteries with ultralow-field nuclear magnetic resonance},
   volume = {17},
   url = {https://pubs.rsc.org/sc/article/17/12/5877-5887/910744},
   year = {2026}
}

@article{Myers2024,
  title = {Zero to ultralow magnetic field NMR of $[1{\text{\ensuremath{-}}}^{13}\mathrm{C}]\mathrm{pyruvate}$ and $[2{\text{\ensuremath{-}}}^{13}\mathrm{C}]\mathrm{pyruvate}$ enabled by SQUID sensors and hyperpolarization},
  author = {Myers, John Z. and Bullinger, Friedemann and Kempf, Nicolas and Plaumann, Markus and Ortmeier, Adam and Theis, Thomas and Povolni, Pavel and Romanowski, Jannik and Engelmann, J\"orn and Scheffler, Klaus and H\"ovener, Jan-Bernd and Buckenmaier, Kai and K\"orber, Rainer and Pravdivtsev, Andrey N.},
  journal = {Phys. Rev. B},
  volume = {109},
  issue = {18},
  pages = {184443},
  numpages = {16},
  year = {2024},
  month = {May},
  publisher = {American Physical Society},
  doi = {10.1103/PhysRevB.109.184443},
  url = {https://link.aps.org/doi/10.1103/PhysRevB.109.184443}
}

\end{document}